\documentclass[twocolumn,apj,numberedappendix,appendixfloats,twocolappendix]{openjournal}

\usepackage{newtxtext,newtxmath}

\usepackage[T1]{fontenc}

\usepackage{graphicx}
\usepackage[colorlinks,linkcolor=blue,citecolor=blue,urlcolor=blue ]{hyperref}
\usepackage{amsmath}
\usepackage{soul}
\usepackage{xspace}
\usepackage{ulem}                   

\usepackage{orcidlink}

\makeatletter 
  
  \patchcmd{\NAT@citex}
    {\@citea\NAT@hyper@{%
      \NAT@nmfmt{\NAT@nm}%
      \hyper@natlinkbreak{\NAT@aysep\NAT@spacechar}{\@citeb\@extra@b@citeb}%
      \NAT@date}}
    {\@citea\NAT@nmfmt{\NAT@nm}%
    \NAT@aysep\NAT@spacechar\NAT@hyper@{\NAT@date}}{}{}

  \patchcmd{\NAT@citex}
    {\@citea\NAT@hyper@{%
      \NAT@nmfmt{\NAT@nm}%
      \hyper@natlinkbreak{\NAT@spacechar\NAT@@open\if*#1*\else#1\NAT@spacechar\fi}%
        {\@citeb\@extra@b@citeb}%
      \NAT@date}}
    {\@citea\NAT@nmfmt{\NAT@nm}%
    \NAT@spacechar\NAT@@open\if*#1*\else#1\NAT@spacechar\fi\NAT@hyper@{\NAT@date}}
    {}{}
\makeatother

\DeclareRobustCommand{\ion}[2]{%
\relax\ifmmode
\ifx\testbx\f@series
{\mathbf{#1\,\mathsc{#2}}}\else
{\mathrm{#1\,\mathsc{#2}}}\fi
\else\textup{#1\,{\mdseries\textsc{#2}}}%
\fi}

\newcommand{\hMsol}{h^{-1}\,{\rm M_\odot}}
\newcommand{\Msun}{{\rm M_\odot}}

\newcommand{\hMpc}{h^{-1}\,{\rm Mpc}}

\newcommand{\ie}{{i.e.~}}
\newcommand{\eg}{{e.g.~}}

\newcommand{\HI}{\ion{H}{I}\xspace}
\newcommand{\HII}{\ion{H}{II}\xspace}

\newcommand{\Lya}{Ly$\alpha$\xspace}
\newcommand{\taueff}{\tau_\mathrm{eff}}

\newcommand{\fesc}{f_{\rm esc}}
\newcommand{\citenp}[1]{\citeauthor{#1} \citeyear{#1}}

\newcommand{\arepo}{{\sc arepo}\xspace}
\newcommand{\areport}{{\sc arepo-rt}\xspace}

\newcommand{\thesan}         {\textsc{thesan}\xspace}
\newcommand{\thesanone}      {\textsc{thesan-1}\xspace}
\newcommand{\thesantwo}      {\textsc{thesan-2}\xspace}
\newcommand{\thesanwc}       {\textsc{thesan-wc-2}\xspace}
\newcommand{\thesanlow}      {\textsc{thesan-low-2}\xspace}
\newcommand{\thesanhigh}     {\textsc{thesan-high-2}\xspace}

\newcommand{\thesansdao}     {\textsc{thesan-sdao-2}\xspace}

\newcommand{\galacc}{GaL$\alpha$CC\xspace} 
\newcommand{\taun}{$\tau_\mathrm{los} - n_\mathrm{gal}$\xspace}
\newcommand{\taulos}{\tau_\mathrm{los}}

\newcommand{\rev}[1]{#1}

\graphicspath{{./}{figures/}}

\shorttitle{The galaxy-IGM connection in \thesan, part 3: the IGM opacity -- galaxy density relation}
\shortauthors{Garaldi et al.}

\begin{document}
\title{\textbf{The galaxy-IGM connection in thesan: the physics connecting the IGM Lyman-$\alpha$ opacity and galaxy density in the reionization epoch}\vspace{-1.5cm}}
\author{E.~Garaldi\orcidlink{0000-0002-6021-7020},$^{1,2,3,4,5,*,\dagger}$}
\author{V.~Bellscheidt\orcidlink{0009-0006-1543-9907}$^{6}$}
\author{A.~Smith\orcidlink{0000-0002-2838-9033}$^{8}$}
\author{R.~Kannan\orcidlink{0000-0001-6092-2187}$^{7}$\vspace{0.2cm}}
\thanks{$^*$E-mail: \href{mailto:egaraldi@sissa.it}{egaraldi@sissa.it}}
\thanks{$^\dagger$CANON Fellow}
\affiliation{$^{1}$Kavli IPMU (WPI), UTIAS, The University of Tokyo, Kashiwa, Chiba 277-8583, Japan}
\affiliation{$^{2}$Institute for Fundamental Physics of the Universe, via Beirut 2, 34151 Trieste, Italy}
\affiliation{$^{3}$SISSA - International School for Advanced Studies, Via Bonomea 265, 34136 Trieste, Italy}
\affiliation{$^{4}$INAF, Osservatorio Astronomico di Trieste, Via G. B. Tiepolo 11, I-34131 Trieste, Italy}
\affiliation{$^{5}$Department of Earth and Space Science, Osaka University, Toyonaka, Osaka 560-0043, Japan}
\affiliation{$^{6}$Technical University of Munich, TUM School of Natural Sciences, Physics Department, James-Franck-Strasse 1, 85748 Garching, Germany}
\affiliation{$^{7}$Department of Physics and Astronomy, York University, 4700 Keele Street, Toronto, ON M3J 1P3, Canada}
\affiliation{$^{8}$Department of Physics, The University of Texas at Dallas, Richardson, TX 75080, USA}

\begin{abstract}  
\noindent The relation between the Lyman-$\alpha$ effective optical depth of quasar sightlines ($\taulos$) and the distribution of galaxies around them is an emerging probe of the connection between the first collapsed structures and the IGM properties at the tail end of cosmic reionization. We employ the \thesan simulations to demonstrate that $\taulos$ is most sensitive to galaxies at a redshift-dependent distance, reflecting the growth of ionized regions around sources of photons and in agreement with studies of the galaxy--Lyman-$\alpha$ cross correlation. This is $d \sim 15 \, \hMpc$ at the tail end of reionization. The flagship \thesan run struggles to reproduce the most opaque sightlines as well as those with large galaxy densities, likely as a consequence of its limited volume. We identify a promising region of parameter space to probe with future observations in order to distinguish both the timing and sources of reionization. We present an investigation of the IGM physical conditions around opaque and transparent spectra, revealing that they probe regions that reionized inside-out and outside-in, respectively, and demonstrate that\rev{, for the range of optical depths probed by our simulation,} residual neutral islands at the end of reionization are not required to produce \rev{highly opaque sightlines}, although they facilitate the task. Finally, we investigate the sensitivity of the aforementioned results to the nature of ionizing sources and dark matter. 
\end{abstract}
\maketitle

\section{Introduction}
\label{sec:intro}
The Epoch of Reionization (EoR, \ie the transformation of intergalactic hydrogen from a neutral state to a hot plasma within the first billion years after the Big Bang), represents one of the current frontiers in the study of structure formation. Its relevance, however, is not limited to its role in transforming the inter-galactic medium (IGM) between galaxies. The onset of the EoR is tightly connected to the growth of the first galaxies in the Universe out of primordial density fluctuations, which are simultaneously sources of this process and impacted by it. Additionally, the ionization and concurrent photo-heating of intergalactic gas changes its cooling, and therefore how it is accreted and retained by dwarf galaxies \citep[\eg][]{Katz2019}, potentially driving a pervasive temporary suppression in star formation \cite[][]{Ocvirk+2021, Cain+2024}. 

The last few years witnessed tremendous progress in the study of the EoR. This was driven by two main factors. Observationally, new-generation facilities (the latest of which is the \textit{James Webb Space Telescope}, or JWST) have dramatically extended our reach, granting us eyes on a sizeable fraction of the primeval galaxy population \citep[although with a limited field of view, see \eg][]{ceersI, EIGERII, Harikane+2023, jades, fresco}. On the theoretical front, radiation-hydrodynamical simulations have started to reach the resolution needed to faithfully simulate galaxy populations within volumes comparable to (although still somewhat smaller than) those needed to have converged reionization histories \citep[\ie $\gtrsim 10^6-10^7 \, \mathrm{Mpc}^3$ depending on the target observable,][]{Iliev+2014, Kaur+2020, GnedinMadau_review}. The most relevant examples of such simulations are CoDa \citep{CoDa,CoDaII,CoDa-AMR,CoDaIII}, CROC \citep{CROC} and \thesan \citep{Thesan_intro, Thesan_igm, Thesan_data, Thesan_Lya}.

Advancements on these two fronts have produced an overall-coherent picture, although many of the details are still escaping our understanding. The ionising photons budget appears to have been dominated by \rev{relatively} small galaxies, thanks to their overwhelming number compared to rarer, more-massive objects \citep[\eg][]{Bouwens+2012, Finkelstein+2019, Atek+2015, Atek+2024, SPHINX20, Thesan_fesc, Kostyuk+2023}, with quasars (QSOs) playing only a negligible role \citep[\eg][]{Obelisk}. However, alternative models are not completely ruled out \citep[see \eg][]{madau15, Naidu+2022, Madau+2024}. 

Reionization appears to have been completed at $z \lesssim 6$ \citep[sometimes called `late-reionization' model, \eg][]{Becker+2015, Bosman+2018, Bosman+2021, Eilers+2018, Zhu+2020, Zhu+2021, Kulkarni2019, Keating+2020, Nasir&DAloisio2020}, with sparse neutral island potentially surviving down to $z \sim 5.3$ \citep{Keating+2020,Nasir&DAloisio2020,Becker+2024}. The evidence for this late reionization comes primarily from observations of the evolution of the Lyman-$\alpha$ (\Lya) forest flux detected in QSO spectra \citep[\eg][]{Fan+2006,McGreer+2011,Yang+2020,Lu+2020,XQR30} and the so-called dark-pixel statistic \citep[\eg][]{McGreer+2011,McGreer+2015,Lu+2020}. The intermediate phases of reionization (\ie $0.1 \lesssim x_\mathrm{HI} \lesssim 1$) are now starting to be constrained using the damping wing \citep{Miralda-Escude1998} of QSOs \citep[\eg][]{Mortlock+2011, Greig+2017,Greig+2019,Greig+2022, Banados+2018, Davies+2018, Yang+2020, Durovcikova+2024, Becker+2024}, galaxies \citep[\eg][\rev{but see \citealt{Keating+2024} for potential limitations}]{Fujimoto+2023, Hayes&Scarlata2023, Hsiao+2023, Jung+2024, Umeda+2024} and --~in a statistical sense~-- of the \Lya forest \citep{Spina+2024, Zhu+2024}. 

The debate on the existence of late-time ($z \lesssim 5.5$) neutral islands in the IGM is ongoing. In fact, three classes of explanations have been put forward to explain the fluctuations in the \Lya forest optical depth ($\tau_\alpha$), all revolving around the fact that \Lya radiation is completely absorbed in gas with \rev{local} neutral fraction $f_\mathrm{HI} \gtrsim 10^{-4}$, while neutral regions are typically defined as those with $f_\mathrm{HI} \gtrsim 10^{-2}$--$10^{-1}$. Therefore, there is a range of $x_\mathrm{HI}$ that qualifies the gas as ionized but still completely absorbs incoming \Lya radiation. 

Quantitatively, \rev{for gas in photoionization equilibrium} $\tau_\alpha$ depends on the \HI photoionization rate ($\Gamma_\mathrm{HI}$), the gas density ($\rho_\mathrm{gas}$) through the baryon overdensity ($\Delta_\mathrm{b}$) \rev{and the temperature-density relation power-law index ($\gamma$), as well as} the gas temperature ($T_\mathrm{gas}$) as \citep[\eg][]{Becker+2015review}:
\begin{equation}
\tau_\alpha \simeq 11 \Delta_\mathrm{b}^{2-0.72(\gamma-1)} \left( \frac{\Gamma_\mathrm{HI}}{10^{-12} \mathrm{s}^{-1}} \right)^{-1} \left( \frac{T_\mathrm{gas}}{10^4 \mathrm{K}} \right)^{-0.72} \left( \frac{1+z}{7} \right)^{4.5}.
\end{equation}
Therefore, fluctuations in $\tau_\alpha$ can be caused by:
\begin{itemize}
	\item \textbf{UVBG fluctuations} \citep{Davies&Furlanetto2016, Nasir&DAloisio2020}. Local variations in the UV background (UVBG) in a fully-ionized Universe can originate from the proximity to the sources of ionizing photons or from a locally-varying mean free path (\eg due to fluctuations in the number density of radiation sinks). These modify the local $\Gamma_\mathrm{HI}$ and therefore the local \HI fraction. Regions of below-average $\Gamma_\mathrm{HI}$ can have sufficient neutral hydrogen to completely absorb the \Lya radiation while remaining fully ionized. 
	\item \textbf{Temperature fluctuations} in the IGM on large-scales \citep{DAloisio+2015}, due to the unequal reionization time of different patches of the IGM. Regions that were recently reionized are hotter than those experiencing an earlier reionization, since they had less time to cool after being photo-heated (it takes approximately $1$--$2$ Gyr to reach an homogeneous IGM temperature after the end of reionization, \eg \citealt{Upton-Sanderbeck+2016}) and therefore have lower $\tau_\alpha$ for the same ionization state. 
	\item \textbf{Residual neutral island} due to incomplete reionization \citep{Kulkarni2019, Keating+2020,Nasir&DAloisio2020}, that completely absorb incoming radiation (both \Lya and ionizing). Recently, \citet{Becker+2024} presented the first direct evidence of the existence of a neutral island at $z<6$. However, it is associated to the most extreme Gunn-Peterson troughs known at such redshift, and therefore it remains unclear whether this implies the existence of such neutral island in less biased regions of the Universe \citep[in fact, troughs like this are expected to be very rare in the standard reionization scenario, ][]{Keating+2020}. \rev{It should be noted that this model also inherently produced strong fluctuations in the UVBG, since within neutral island there is a vanishing number of ionising phtons. These fluctuations are much stronger than in the `UVBG fluctuations' model aforementioned. The difference between these two models resides in the presence or absence of large neutral patches in the IGM associated with \Lya absorption, opposed to smoother environment fluctuations. }
\end{itemize}

These different models yield somewhat different predictions for the relationship between sightline opacity and galaxy density around it (\taun relation hereafter). In the UVBG fluctuations model, the most opaque (to \Lya photons) sightlines are associated with underdense regions. Since fewer sources are present, the ionising photon density is suppressed, and so is the \HII fraction (that however remains close to unity), increasing the (average) sightline opacity. For temperature fluctuation models, instead, opaque sightlines are associated to regions that reionized earlier than average, which tend to be overdense since galaxies reside preferentially in such regions. In both cases, transmissive sightlines behaves in an opposite way to opaque ones. Finally, in models with residual neutral highlands there is a large variety of galaxy densities associated to transmissive lines of sight, although they appear to be preferentially associated with overdensities \citep{Nasir&DAloisio2020}. In a realistic scenario, fluctuations in both UVBG and gas temperature are present, and potentially residual neutral islands as well.

\rev{Thanks to its smaller oscillator strength, the Lyman-$\beta$ transition can be used to probe regions of higher densities and/or lower UV radiation field, potentially delivering information on the state of the gas in regions of high $\tau_\alpha$. However, the contamination from the foreground \Lya forest practically inhibits an investigation of individual IGM regions in most cases, leaving only statistical studies averaging over large samples. A possible way to circumvent this has been recently explored in \citet{Meyer+2025}, where galaxies have been used as background sources instead of bright quasars. This allows to easily find pairs of galaxies almost aligned on the sky but at different redshifts, and therefore to use the \Lya forest in the foreground source to clean the background spectrum from its contamination and recover the Lyman-$\beta$ signal. However, the accuracy of available observations is not yet sufficient to fully exploit this method.}

Observations of the \taun relation have slowly started to become available in the last few years, although only for a handful of sightlines \citep{Becker+2018, Kashino+2020, Christenson+2021, Christenson+2023, Ishimoto+2022}. These authors have investigated a total of 7 sightlines to $z > 6$ QSOs, including some of the most transparent and most opaque known. From this limited sample, it appears that the most transparent ($\tau_\mathrm{los} \lesssim 2.5$) \textit{and} most opaque ($\tau_\mathrm{los} \gtrsim 5.5$) sightlines show an underdensity of galaxies within $10 \, \hMpc$, while sightlines with intermediate transmissivity  ($4 \lesssim \tau_\mathrm{los} \lesssim 5$) show an overdensity \citep{Christenson+2023}. 
Intriguingly, none of the theoretical models described above can fully predict the variation in the surface density of LAEs as a function of distance from the line of sight for all observed quasar fields. 

While most of these observations used \Lya emitters (LAEs) as tracers of the underlying density distribution, \citet{Kashino+2020} \rev{obtained equivalent results} employing Lyman break galaxies (LBGs) for the same field observed by \citet{Becker+2018}. \rev{This renders} less likely the possibility that overdensities estimated using LAEs are not a good proxy of the real ones because of the physics involved in the production and escape of \Lya radiation. \rev{However studies of (different) quasar environments (opposed to the environment around the sightline being discussed here, but conceptually similar) showed that LAEs and LBGs can trace different structures. For instance, \citet{Goto+2017} and \citet{Ota+2018} found that underdensities of LAEs corresponded to overdensities of LBGs.}

In this paper, we employ the \thesan radiation-hydrodynamical simulation suite to provide predictions concerning the \taun relation and compare them to available observations. We also exploit the fact that \thesan provides different physical models to explore how this quantity depends on the population of galaxies driving reionization as well as on the nature of dark matter itself. 
The paper builds on top of a companion paper (Garaldi \& Bellscheidt, submitted, hereafter GB24) where we explore the galaxy-\Lya cross-correlation (\galacc), which encodes information on the interplay between the galaxy population and the IGM during cosmic reionization. The manuscript is organised as follows. In Sec.~\ref{sec:methods} we described the simulation suite used and the production of synthetic \Lya forest spectra. In Sec.~\ref{sec:results} and Sec.~\ref{sec:results_variations} we present our predictions regarding the \taun relation in the flagship \thesanone simulation and in those with altered physical models, respectively. Finally, we discuss our findings and provide concluding remarks in Sec.~\ref{sec:conclusions}. All quantities are in comoving units throughout the paper.

\section{Methods}
\label{sec:methods}
Reionization is a cosmological process that requires large volumes to be studied. At the same time, global galaxy properties demand resolutions of order $\mathcal{O}(100 \,\mathrm{pc})$ to be simulated. These simultaneous requirements are challenging to meet, but mandatory in order to faithfully investigate the galaxy-reionization interplay. In this paper, we employ one of the few (see Sec.~\ref{sec:intro}) simulations able to meet these requirements --~albeit marginally, especially in the case of the non-flagship runs~-- namely \thesan. This simulation suite is thoroughly described in \citet{Thesan_intro} and \citet{Thesan_data}, but we provide a brief summary below for the sake of completeness.

\begin{figure*}
    \includegraphics[trim={0 0 1.8cm 0},clip,width=\textwidth]{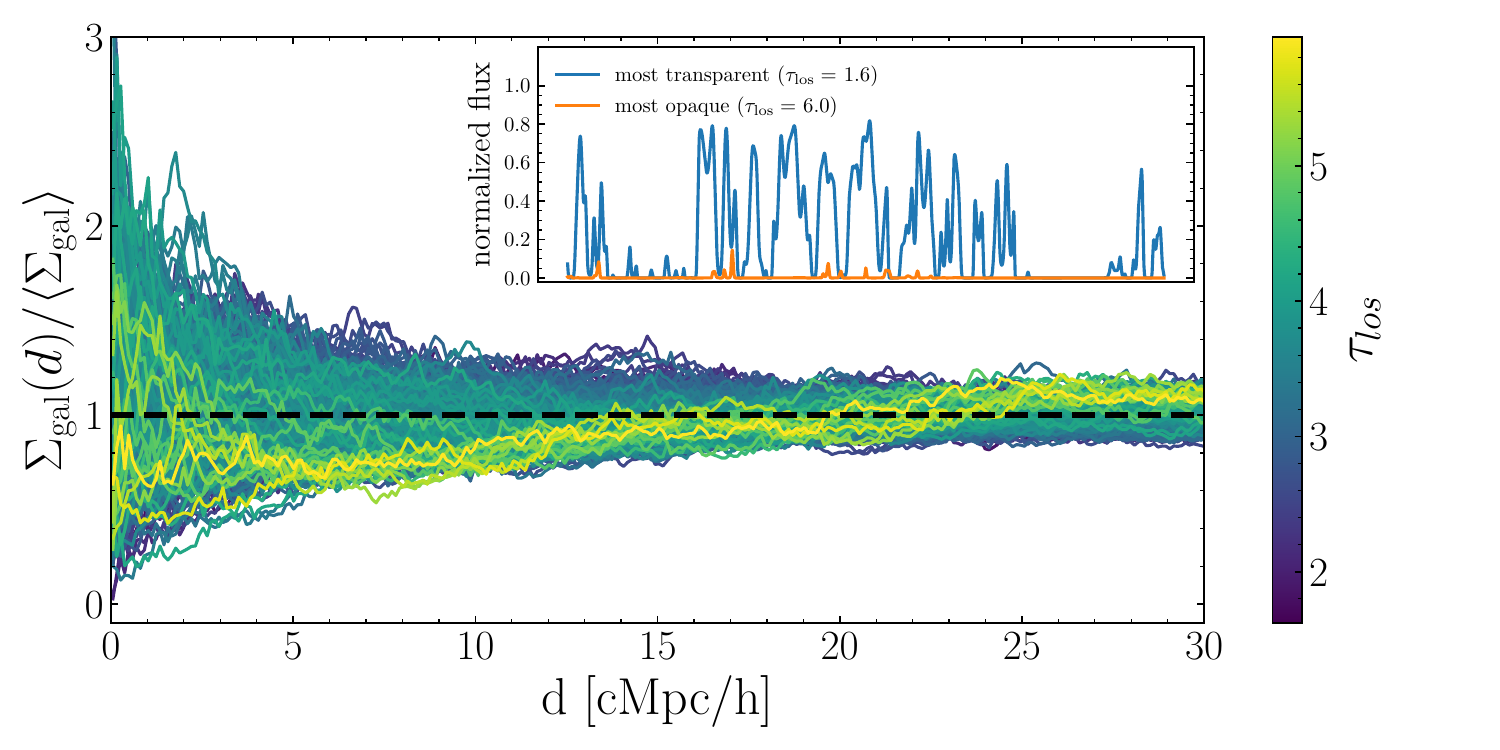}
    \flushright \includegraphics[trim={2cm 1.25cm 1.5cm 6cm},clip,width=1\textwidth]{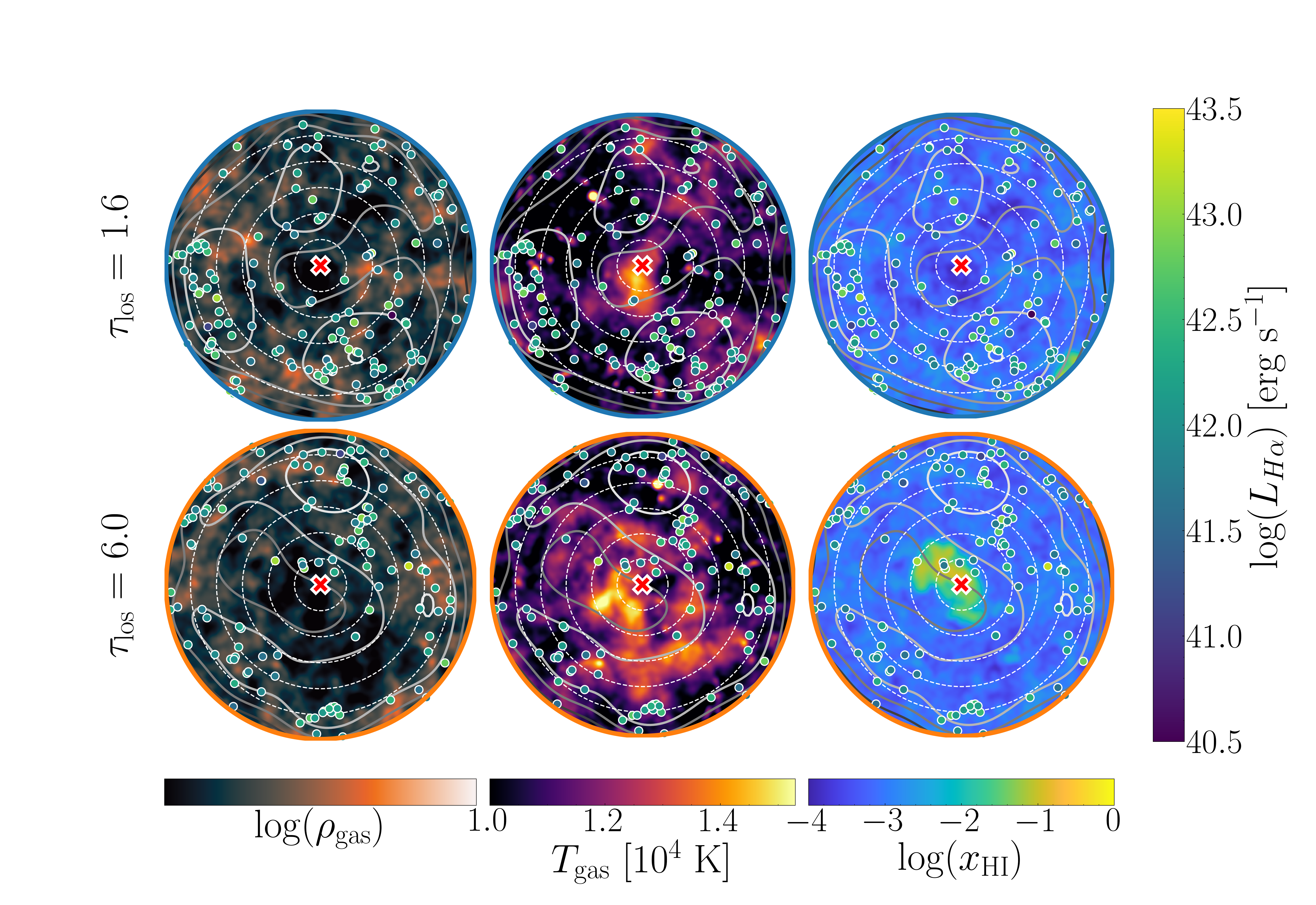}
    \caption{\textbf{Top}: Galaxy overdensity as a function of distance from each of the 600 lines of sight extracted at $z=5.7$ from the \thesanone simulation, color-coded by their \Lya optical depth ($\taulos$). The black dashed line indicate the average overdensity in the simulation (\ie 1 by definition) and is used to guide the eye. 
    The inset shows the normalized \Lya flux in the most transparent (blue line) and most opaque (orange line) lines of sight. \textbf{Middle and bottom}: Distribution of galaxies and gas properties around two simulated sightlines. The sightline position is indicated by a red cross, and its direction (perpendicular to the plane of the figure) is the projection axis. Circles indicated the projected galaxy position and are color-coded by the galaxy H$\alpha$ luminosity. The projected density of \textit{all} galaxies (reconstructed using a Gaussian kernel density estimator) is shown by the contours. The background maps show the average gas density (left), temperature (centre) and \HI fraction (right) along the projection axis. We also plot circles with radius increasing by steps of 5 Mpc, up to 30 Mpc. The middle row shows the most transparent sightline in the simulation ($\taulos = 1.6$), while the bottom one refers to the most opaque line of sight ($\taueff = 6.0$). 
    }
    \label{fig:ngal_R_visual}
\end{figure*}

\subsection{The \thesan simulations}
\label{subsec:thesan}
\thesan is a suite of radiation-hydrodynamical simulations built with the explicit goal of bridging the gap between reionization and galaxy formation studies. It has been recently made publicly available at \url{www.thesan-project.com} \citep{Thesan_data} and is able to capture the observed galaxy \citep{Thesan_intro, Thesan_data, Thesan_sizes} and IGM properties \citep{Thesan_igm}. At the same time, it employs a combination of models calibrated at $z\lesssim 4$ in order to have a single free parameter, namely the escape fraction ($\fesc$) of ionizing radiation from the birth loci of stars\footnote{Notice that this is \textit{not} the `escape fraction' that enters in reionization models, which instead represents the escape from an \textit{entire galaxy/halo}. The latter can be self-consistently predicted by \thesan \rev{(given its resolution and physical model)}, as done in \citet{Thesan_fesc}.}, that we loosely calibrate requiring \thesan to follow a `late' reionization history completing at $z \lesssim 5.5$. More specifically, \thesan employs the successful IllustrisTNG model for galaxy formation \citep{Weinberger2017, Pillepich2018b}, the cosmic dust model of \citet{McKinnon+16} and the \areport \citep{ArepoRT} radiation transport solver embedded in the \arepo code \citep{arepo,Arepo-public}. 
\rev{This uses a moment method with M1 closure \citep{Levermore84} to solve the radiation-transport equations. This is known to produce a smoother-than-expected radiation field and to degrade in accuracy in the post-reionization universe, while yielding the expected results during the EoR \citep{Wu+2021}. Additionally, \thesan employs a reduced-speed-of-light approximation to ease the computational demands of fully-coupled radiation-hydrodynamics. This has been calibrated to ensure that the simulated reionization history is not affected by this numerical parameter. However, \citet{Cain+2024} recently showed that this might still affect the simulated \Lya forest, although additional investigation is required since these authors used a very different (spatially less accurate, directionally more precise) method to solve the radiation transport equations.}

The cosmology employed in all \thesan simulations is the \citet{Planck2015cosmo} one. The \thesan suite includes a number of different simulations, all following a cubic region of the Universe with linear size $L_\mathrm{box} = 95.5 \, \mathrm{Mpc}$. The flagship run, \thesanone, has the highest resolution, sufficient to resolve atomic cooling haloes and therefore to capture the vast majority of the ionizing photons budget. This run is flanked by a range of simulations with different physical models. These have mass resolution decreased by a factor of $8$, and their $\fesc$ has been re-caibrated to account for the missing photons from the unresolved atomic cooling haloes (except \thesantwo, which is used for numerical convergence studies). 

\rev{In the fiducial \thesan model, $\fesc$ is a universal value. Therefore, we do not expect this parameter to change the relative behaviour of different classes of sightlines discussed in this paper, nor their relation to the local galaxy distribution, as long as they belong to the same simulation box. Only in the extreme case of extremely early or delayed reionization history we might expect a significant impact of this parameter on our results.}

Although \thesan is among the largest-volume radiation-hydrodynamical simulations of the Universe able to capture galaxy properties (alongside CROC --~\citealt{CROC}~-- and CoDaIII --~\citealt{CoDaIII}), its volume is still too small for a direct and faithful comparison with observations spanning tens (if not hundreds) of Mpc in each dimension as the one addressed in this paper. For instance, the observations in \citet{Ishimoto+2022} and \citet{Christenson+2023} probe radii as large as $\approx 70 \, \hMpc$, which are \textit{larger} than the linear size of our simulation box. 
Conversely, the largest radial distance that can be probed by our simulation is $d_\mathrm{max} \sim L_\mathrm{box}/2 \approx 32 \, \hMpc$, since the periodicity of the simulation box implies that the same structures are re-sampled at larger separations. 
Nevertheless, we believe that the results in this paper remain interesting for a number of reasons, namely: (i) we provide predictions that, unlike those available for the \taun relation until now, stem from a state-of-the-art radiation-hydrodynamical simulation that was explicitly designed to investigate the interplay between IGM and galaxies, therefore reaching the highest degree of physical fidelity among the different approaches to the modeling of the early Universe; (ii) we provide predictions for the \taun relation under different assumption for some of the uncertain physical processes (including the galaxy population powering reionization and the nature of dark matter); and finally, (iii) as we will show below, as well as from the results presented in GB24, the most relevant scales for the \taun relation are of order $d \lesssim 25 \, \hMpc$, well within the range of scales that \thesan can faithfully probe without incurring in artifacts due to the finite volume. 

\rev{Finally, we note that recent work showed that it is possible that scales below our resolution affect the propagation of ionising photons \citep{DAloisio+2019, Cain+2023, Cain+2024}. The extent of such effect in our simulation is difficult to estimate, given the different methods employed, but will be studied in a future work.}

\subsection{Synthetic spectra}
\label{subsec:spectra}
In this paper we make use of synthetic lines of sights (LOS). These were extracted from the simulation outputs as detailed in Section 3.10 of \citet{Thesan_data}. In short, we use the \textsc{colt} code \citep[last described in][]{Smith+2022}, which is able to retain the full spatial information by directly exploiting the native Voronoi tessellation of \arepo, to extract gas properties along 600 independent LOS, each with length $50$ Mpc, for each simulation snapshot (\ie approximately every 11 Myr) between $5.5\lesssim z \lesssim 7$. We employ the same window to identify galaxies round the LOS. 
These are somewhat shorter than those used in \citet{Ishimoto+2022} and \citet{Christenson+2023}, \ie $50 \, h^{-1}$ cMpc. We show in Appendix~\ref{app:different_spectral_lengths} that this does not affect our conclusions, and concurrently discuss the impact of identifying galaxies only in the central part of the spectrum as a consequence of the filter used for this goal.
In the construction of the synthetic \Lya forest spectra we use the approximation of \citenp{Harris1948} and \citenp{Tepper-Garcia2006} to the full Voigt-Hjerting line profile \citep{Hjerting1938}, including temperature broadening of the line and the gas peculiar velocities. The native spectral resolution of our spectra is $\Delta \varv = 1 \, \mathrm{km\,s}^{-1}$. 
\rev{Finally, mirroring what is typically done in observations, we associate to each sightline an effective optical depth $\taulos = -\log (\langle f_\mathrm{Ly\alpha} \rangle_\mathrm{los})$, where $\langle f_\mathrm{Ly\alpha} \rangle_\mathrm{los}$ is the mean transmitted (normalised) \Lya flux along the line of sight.}

\section{Results from thesan-1}
\label{sec:results}
In this section, we present the predictions from the flagship \thesanone simulation. We focus on redshift $z=5.7$ because all available observations are centered at this time \citep[][]{Becker+2018, Kashino+2020, Christenson+2021, Christenson+2023, Ishimoto+2022}. Therefore, unless otherwise stated, all plots refer to this redshift. In the next Section, we will present the results from other runs from the same suite exploring physical variations. 

\subsection{Radial sensitivity}
The study of the \galacc indicates that, on average, the IGM is maximally impacted by the radiation field of a galaxy at a specific distance scale, which evolves with redshift and traces the progress of reionization \citep{Thesan_igm}. 
This results from the competition between ionizing radiation field of the galaxy and the overdensity where these galaxy reside. The latter boosts recombination and, therefore, acts against the former. Both effects grow in strength when approaching the galaxy, but their radial dependence is different. In particular, the influence of the galactic radiation field extends much farther than the local overdensity (at least after the initial stages of reionization). This difference creates the peculiar radial modulation of this effect (GB24). For the late reionization history simulated in the \thesan suite, at $z=5.7$ 
this distance corresponds to approximately $15 \, \hMpc$. 
It is therefore a plausible hypothesis that the opacity of a sightline is mostly affected by galaxies located at such specific distance from the line of sight. 
\rev{We note that a multitude of effects can break this relation for individual galaxies and/or sightlines (\eg self-shielded clumps, variation in the neutral gas content of haloes, etc.). However, when averaged over sufficiently large samples, such `structure formation noise' contributes merely to the scatter around the mean relation. We have shown this explicitly (in the context of the \galacc) in GB24.}

We begin our investigation of the aforementioned hypothesis through the visualisation in the top panel of Fig.~\ref{fig:ngal_R_visual}, where we plot the projected galaxy overdensity ($\Sigma_\mathrm{gal} (d) / \langle \Sigma_\mathrm{gal} \rangle$\rev{, computed over the entire sightline, \ie $50\,\mathrm{cMpc}$, see Sec.~\ref{subsec:spectra}}) as function of distance ($d$) from each of the 600 LOS used in this work. The color of each line indicates the optical depth ($\taulos$) of the quasar LOS (as reported in the right-hand-side colorbar). Lines are drawn starting from the most transparent and ending with the most opaque, in order to ease the visualization of the latter. 
To better show the variability in transmissivity that characterizes the ending phases of reionization, we plot in the inset panel the normalized \Lya flux in the most transparent ($\taulos=1.6$) and most opaque ($\taulos=6.0$) sightlines in \thesanone. We also show a more detailed view of these two sightlines in, respectively, the middle and bottom rows of Fig.~\ref{fig:ngal_R_visual}. Each panel in these rows shows a projections centred on the sightline (indicated by red cross) and along its direction. The background maps show the average gas density (left), temperature (centre) and \HI fraction (right), while galaxies with mass $M_\mathrm{star} \geq 5 \times 10^8 \, \hMsol$ are indicated by circles, color-coded by the galaxy H$\alpha$ emissivity \citet[estimated from its star-formation rate following][]{Kennicutt+1994}. Finally, the contours show the projected density distribution reconstructed from \textit{all} galaxies using a Gaussian kernel density estimator, while dashed lines indicate the distance from the sightline in steps of 5 Mpc. 

\begin{figure}
    \includegraphics[width=\columnwidth]{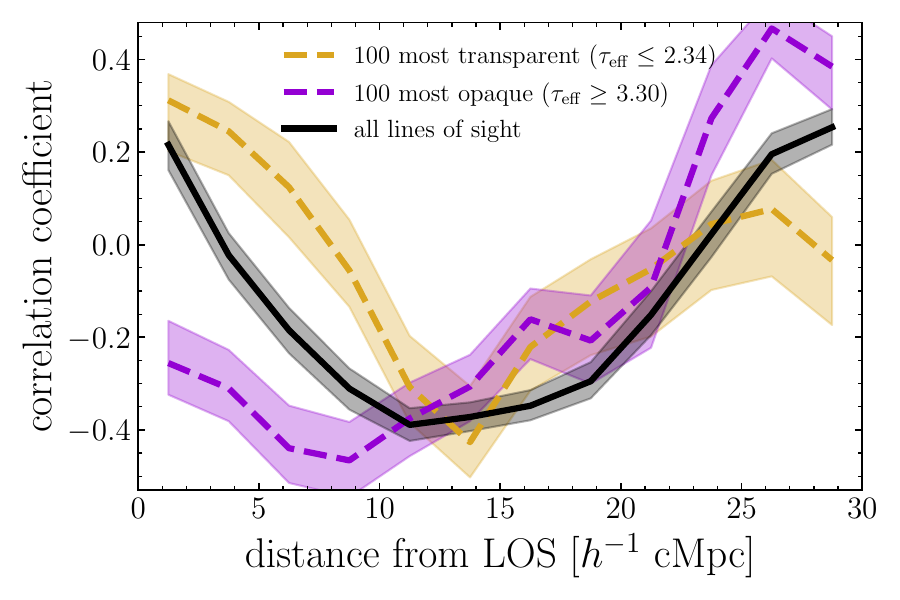}
    \caption{Pearson correlation coefficient between the number of galaxies in a cylindrical annulus around the LOS and the total Lyman-$\alpha$ optical depth in the LOS, as function of the annulus radius. \rev{The Figure shows that the LOS \Lya optical depth is maximally sensitive to galaxies at distance $\approx 15 \, \hMpc$ from it.}}
    \label{fig:ngt_corrcoeff}
\end{figure}

The top panel of Fig.~\ref{fig:ngal_R_visual} visually shows that at both small and large separations, the most transparent and opaque LOS show number densities of galaxies in line with less extreme ones. At distances $5 \lesssim d/[\hMpc] \lesssim 15$, however, the situation is different. It appears that opaque (transparent) sightlines tend to exhibit a lower (higher) number density of galaxies at such distances with respect to the mean in the simulation (horizontal black dashed line). 
In order to quantitatively assess this behaviour, we compute the Pearson correlation coefficient between the values of $\tau_\mathrm{los}$ and of $n_\mathrm{gal}$ computed in radial bins of width $\Delta d = 3 \, \hMpc$. We present the results in Fig.~\ref{fig:ngt_corrcoeff}, where the correlation coefficient is plotted as function of the (central) bin radius for three different samples, namely: all LOS (black solid line), the 100 most transparent ones (dashed yellow line, corresponding to $\taulos \leq 2.34$) and the 100 most opaque ones (dashed purple line, corresponding to $\taulos \geq 3.30$). For each of them, we also report with a shaded region the central 68\% of the distribution of correlation coefficient obtained through bootstrapping. While the existence of a correlation is clear, the correlation coefficient is moderate, indicating that a substantial amount of variability is to be expected, thus underscoring the importance of increasing the sample size of this kind of observations, that are currently limited to merely 7 lines of sight.

The solid black line in Fig.~\ref{fig:ngt_corrcoeff} clearly shows that, when considering all LOS, the opacity is most sensitive to galaxies within $10 \lesssim d/[\hMpc] \lesssim 20$, as hypothesized above. 
It is interesting to note that there seems to be a positive correlation of $\taulos$ with the number density of galaxies at very low and very large distances. While the former is easily understood as the effect of the galaxy overdensity boosting hydrogen recombination and therefore opacity, the latter is puzzling. In fact, we do not expect galaxies at such distance to affect the LOS properties because of their distance. 
\rev{We have carefully checked that the correlation vanishes when any of the samples is randomly re-shuffled, as well as that the outcome is not affected by the choice of radial bins, the galaxy selection, or the size of the `most opaque' and `most transparent' samples. While we have been unable to identify unambiguous explanation for the large-scale behaviour, we notice that the most opaque sightlines tend to be less dense at distances $d \lesssim 10\,\hMpc$ compared to the rest of the sample. 
This can create a positive correlation at $d \gtrsim 20\,\hMpc$ due to one or a combination of the following reasons. 
First, by virtue of the fact that the box has the average Universe density, the lack of galaxies at intermediate distances must be compensated by an excess at larger scales, therefore creating an (unphysical) positive correlation. 
Second, the void size distribution at $z=5.7$ is predicted to peak at void radius $R_\mathrm{void} ~ 15 \, \hMpc$ \citep{DAloisio&Furlanetto2007}. 
Using this as a proxy for the typical distance between an underdensity and an overdenity of galaxies means that the fewer galaxies at $d \sim 10\,\hMpc$ imply more of them at $d \sim 25\,\hMpc$. This implies a physical but indirect positive correlation at large scales only for very opaque sightlines. 
The effect described is driven by the most opaque sightlines but affects the full sample because (i) it is present for approximately 50\% of the LOS investigated and (ii) it is not compensated by an opposite trend in transparent sightlines.}

When restricting our analysis to the 100 most transparent sightlines, we find similar results as for the entire sample, although with a somewhat narrower minimum and without the puzzling positive correlation at large distances. Finally, the sample containing the 100 most opaque sightlines shows a different behaviour. The positive correlation at $d \sim 0$ is not present, while the one at the largest distances is even stronger than for the full sample. Finally, the maximum anticorrelation is shifted to smaller radii ($d \sim 7 \, \hMpc$). We will discuss the physical reason for these differences in Sec.~\ref{sec:physical_properties}. 

\begin{figure}
    \includegraphics[width=\columnwidth]{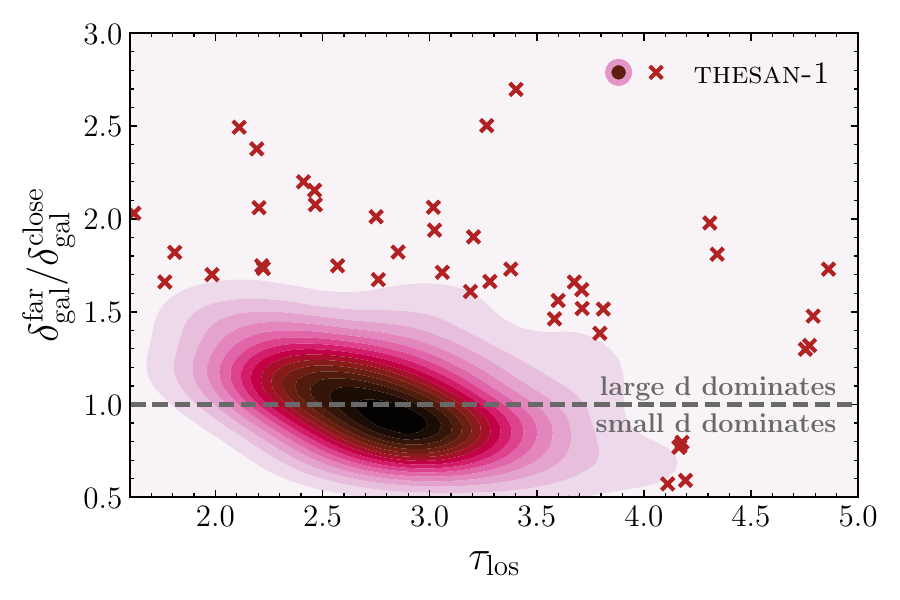}
    \caption{Two-dimensional distribution of large-to-small-scale-overdensity ratio ($\delta_\mathrm{gal}^\mathrm{far} / \delta_\mathrm{gal}^\mathrm{close}$) and optical depth ($\taulos$) for each of the investigated lines of sight. The color reflects the number density of sightlines estimated using a Gaussian kernel density estimator, while crosses show individual sightlines in regions where the estimated density is below 5\% of the maximum. }
    \label{fig:oden_ratio_vs_tau}
\end{figure}

To further test the underlying hypothesis to our explanation, \ie that galaxies within $\sim 5 \, \hMpc$ increase the LOS opacity and those at $d \sim 15 \, \hMpc$ instead decrease it, we compute for each sightline the galaxy overdensity within $7 \, \hMpc$ ($\delta_\mathrm{gal}^\mathrm{close}$) and between $7 \, \hMpc$ and $30 \, \hMpc$ ($\delta_\mathrm{gal}^\mathrm{far}$), taken as representative of the impact of nearby and far galaxies. 
In Fig.~\ref{fig:oden_ratio_vs_tau} we plot the two-dimensional distribution of $\taulos$ and $\delta_\mathrm{gal}^\mathrm{far}/\delta_\mathrm{gal}^\mathrm{close}$, estimated using a Gaussian kernel density estimator (and showing with crosses the location of sightlines in regions where the estimated density is below 5\% of the maximum). The ratio on the vertical axis is a measure of the relative importance of the galaxy density far and close to the galaxy. There is a clear anticorrelation between the value of this ratio and the opacity of the sightline. We conclude that the same physical processes responsible for the \galacc are at work here, shaping the \taun relation. Therefore, we expect a strong redshift evolution (driven by the evolving IGM neutral fraction) of the curve shown in Fig.~\ref{fig:ngt_corrcoeff}, that we investigate next.

\subsubsection{Redshift dependence}
\label{subsec:redshift_dependence}

\begin{figure}
    \includegraphics[width=\columnwidth]{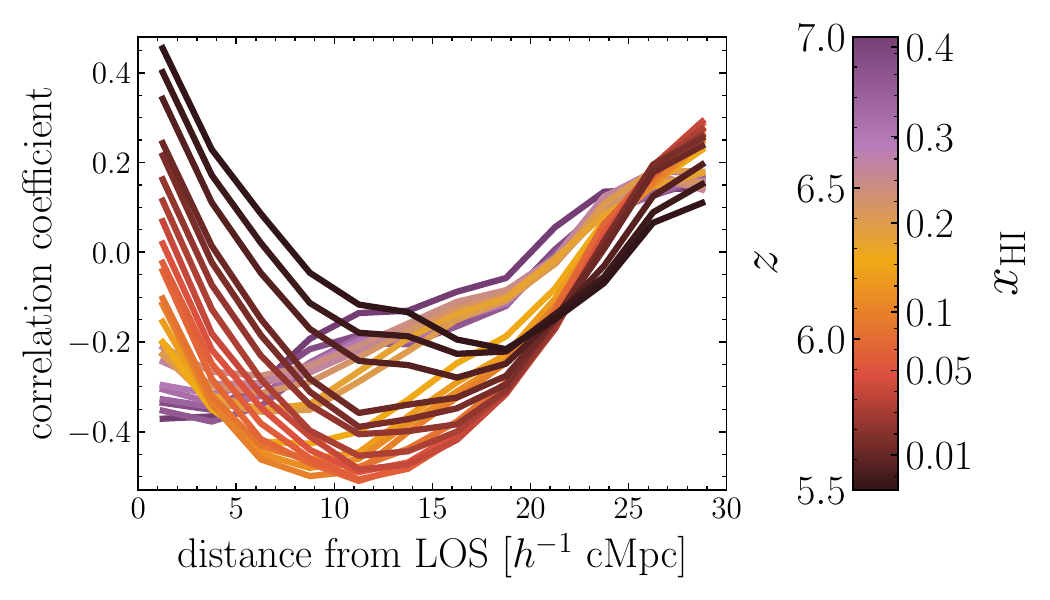}
    \caption{As Fig.~\ref{fig:ngt_corrcoeff} but only showing the curve including all sightlines and for different redshifts (\ie different values of the volume-averaged hydrogen neutral fraction $x_\mathrm{HI}$).}
    \label{fig:ngt_corrcoeff_evol}
\end{figure}

Drawing upon the tight connection between the \taun and \galacc relations depicted above, we expect the scale of the maximum anti-correlation between the galaxy overdensity and the sightline opacity (as well as its strength) to evolve rapidly with the progress of cosmic reionization. We show this explicitly in Fig.~\ref{fig:ngt_corrcoeff_evol}, where we report the correlation coefficient computed for all sightlines (\ie the black line in Fig.~\ref{fig:ngt_corrcoeff}) as a function of redshift (indicated by the line color). 
The evolution of these curves matches very well the prediction informed by the \galacc. In fact, at higher redshift the sightline \Lya opacity is mostly influenced by very close galaxies, since the spatial extent of their proximity effect grows with time (\eg GB24). This not only further clarifies the connection between the \galacc and the \taun relation, but also offers a clue on the difference observed in Fig.~\ref{fig:ngt_corrcoeff} between opaque and transparent sightlines. The former (latter) are probing regions where the local reionization history is delayed with respect to (ahead of) the global one, and therefore are sensitive to galaxies located at different scales. We have explicitly checked that this is the case in the simulation, although the scatter in this relation is significant.

\subsection{Comparison with observations}
\label{subsec:comparison_with_obs}
Before moving on to the analysis of the physical differences between simulated and opaque sightlines, we perform here a comparison with available data, namely from \citet{Ishimoto+2022} and \citet{Christenson+2023}.\footnote{Note that \citet{Christenson+2023} re-analyses the sightlines from \citet{Becker+2018} and \citet{Kashino+2020}, which therefore are automatically included in our comparison.} Observationally, these studies used LAEs to trace the galaxy distribution. Unfortunately, simultaneously simulating the production and escape of \Lya photons on $\mathcal{O} (100 \mathrm{Mpc})$ scales is beyond the reach of even the most advanced simulations. 
A previous study by \citet{Keating+2020} employed an empirical probabilistic model to assign a \Lya luminosity to simulated galaxies, based on the UV magnitudes. While this approach captures the fact that not all UV-bright galaxies are LAE, it does so at the price of introducing additional free parameters. Therefore, here we opt for a simpler approach. We select for \rev{our} analysis all galaxies above a stellar mass threshold $M_\mathrm{star, thr}$. We fix its value by requiring that the surface density of selected galaxies at large distances from the LOS matches the observed one \citep[\ie $\Sigma_\mathrm{LAE} = 0.02 \, h^2/\mathrm{Mpc}^2$,][]{Christenson+2023}. This approach is corroborated by the finding of \citet{Kashino+2020} that, observationally, results are identical when Lyman Break Galaxies are used in place of LAEs. For \thesanone, the resulting threshold mass is $M_\mathrm{star, thr} \approx 10^9 \, \hMsol$.

\begin{figure}
    \includegraphics[width=\columnwidth]{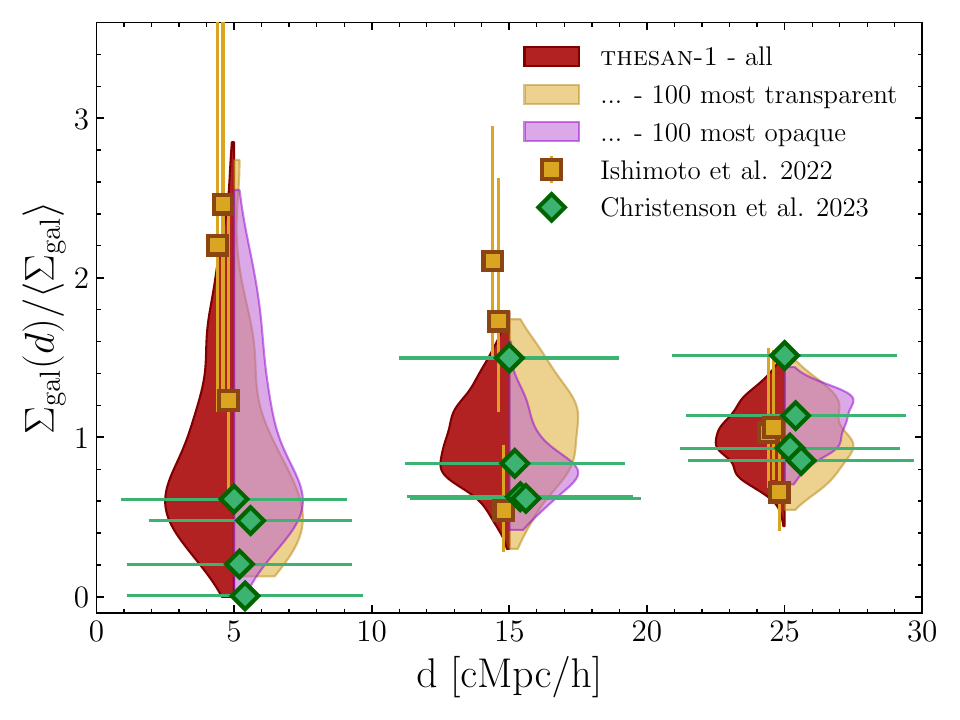}
    \caption{Distribution of galaxy overdensity $\Sigma_\mathrm{gal} (d) / \langle \Sigma_\mathrm{gal} \rangle$ at different radii $d$ in \thesanone, compared to the observed values by \citet{Ishimoto+2022} and \citet{Christenson+2023} (slightly offset in the horizontal direction for visual clarity). The galaxy overdensity is computed as in observations, after selecting galaxies to have the same number density as observed (see text for details on this procedure) and using the same bins in radial distance. The red left-side violins are computed using all sightlines in the simulation, while the yellow and purple right-side violins are computed using only, respectively, the 100 most transparent and most opaque lines of sight. }
    \label{fig:oden_vs_dist_obs}
\end{figure}

After performing the aforementioned selection, we show in Fig.~\ref{fig:oden_vs_dist_obs} the range of overdensities found in the simulated sightlines within the same radial bins used in \citet[][yellow squares]{Ishimoto+2022} and \citet[][green diamonds]{Christenson+2023}. The left-side red violins show the overdensity distribution across all sightlines, while the two right-side violins report the one obtained for the 100 most transparent (yellow violins) and opaque (purple violins) lines of sight. Despite the somewhat small box compared to the scale of observations, \thesanone shows a range of overdensities very similar to the observed one. It should be noted that observations extend to much larger distances, but the limited box size of the simulation prevents us from properly sampling such scales, and we therefore do not show them in the figure. Interestingly, the sightlines in \citet{Ishimoto+2022} were selected to have moderate optical depth and, therefore, be more representative of the average population, but they appear to be more extreme in their overdensities than those observed by \citet{Christenson+2023}, preferentially residing in the tail of the simulated distribution at $d \leq 15 \, \hMpc$. 

Fig.~\ref{fig:oden_vs_dist_obs} clearly shows that transparent sightline preferentially have larger galaxy densities at $d \approx 15 \, \hMpc$. 
\rev{To quantitatively substantiate this claim, we performed a Kolmogorov–Smirnov (KS) test. The null hypothesis that the most transparent and most opaque sightlines are extracted from the same parent distribution is rejected (p-value $< 0.05$) at all distances.}  
At larger distances, there is a small difference in the opposite direction, while at smaller radii the three distributions are essentially identical. This is in line with the conclusion of \citet{Christenson+2023} that underdensities around transparent lines of sight extend for approximately $10 \, \hMpc$, while those around opaque sightlines have scales twice as large.

Additionally, we also find that both the most transparent and the most opaque lines of sight show underdensities at $d \lesssim 10 \, \hMpc$. However, this is not a peculiarity of the sightlines themselves, but rather a general feature. The reason is simply that they probe random regions of the IGM (since the proximity zone of the background quasar is removed from the analysis), and overdensities cover only a small \textit{volume} fraction of the Universe. Therefore, it is much more likely for a sightline to show an overall underdensity of galaxies around it than an overdensity. 
The data from \citet{Ishimoto+2022}, however, appear to probe extreme overdensities \rev{in our simulation} at both intermediate and small distances. Our previous analysis indicates that this could actually be the \textit{reason} for their intermediate opacities. In fact, the large abundance of galaxies at $d \approx 15 \, \hMpc$ boosts the transmissivity, while the galaxies at $d \lesssim 10 \, \hMpc$ suppress it, resulting in intermediate values of $\taulos$. However, this would imply an (un)fortunate coincidence, since the selection was done purely on their effective optical depth, which could be achieved more easily by much more typical environments, as we show next. \rev{One caveat to consider is that the finite volume of our simulations entails that we are unable to capture the largest density fluctuations in the Universe.}

\begin{figure}
    
    \includegraphics[width=\columnwidth]{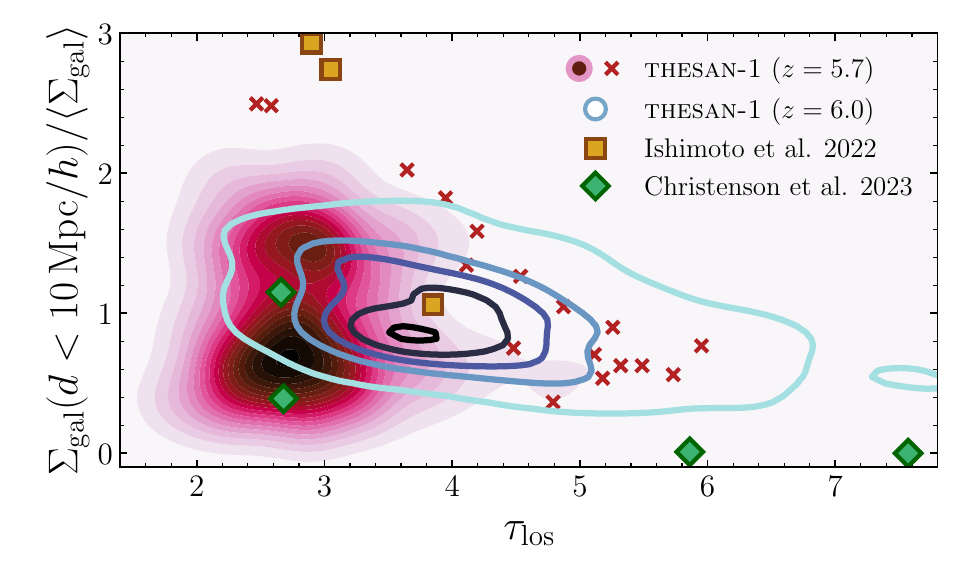}
    \caption{Two-dimensional distribution of galaxy overdensity within 10 Mpc/$h$ of the sightlines ($\Sigma_\mathrm{gal} (d < 10 \, \mathrm{Mpc}/h) / \langle \Sigma_\mathrm{gal} \rangle$) and sightline optical depths ($\taulos$). The filled (empty) contours color reflects the number density of sightlines at $z=5.7$ ($z=6.0$) estimated using a Gaussian kernel density estimator, while crosses show individual sightlines in regions where the estimated density is below 5\% of the maximum (only for $z=5.7$). Diamonds show the measurements from \citet{Ishimoto+2022} and \citet{Christenson+2023}. The galaxy overdensity is computed as in observations, after selecting galaxies to have the same number density as observed (see text for details on this procedure). 
    }
    \label{fig:oden_vs_tau_obs}
\end{figure}

In Fig.~\ref{fig:oden_vs_tau_obs} we show the two-dimensional distribution of $\taulos$ and galaxy overdensity within 10 $\hMpc$ of the sightlines. The filled contours show the estimated distribution in the \thesanone simulation using a Gaussian kernel density estimator (we additionally show individual sightlines in regions where the estimated density is below 5\% of the maximum using red crosses). The observations from \citet{Ishimoto+2022} are shown using yellow squares and those by \citet{Christenson+2023} using green diamonds. In this case the agreement between observations and simulations is worse than in the previous Figure. In particular, \thesanone struggles to reproduce the large galaxy density found by \citet{Ishimoto+2022} and the very opaque sightlines from \citet{Christenson+2023}. For the latter, however, it should be noted that they were selected as two of the most opaque sightlines known, and are therefore not expected to be found in boxes of somewhat limited volume. Alternatively (or concurrently), the reionization history in \thesanone might still be slightly too early compared to observations \citep[as also concluded in][]{Thesan_igm}. At $z=6$ (corresponding approximately to a shift of $\Delta z = 0.3$ in the reionization history), the simulation is able to reproduce the sightlines with $\taulos \sim 6$--$7$. In the figure this can be seen looking at the empty blue contours, that are identical to the filled ones but computed at $z=6$ (instead of the fiducial $z=5.7$). However, the tension with the points showing $\Sigma_\mathrm{gal} (d < 10 \, \mathrm{Mpc}/h) / \langle \Sigma_\mathrm{gal} \rangle \sim 3$ somewhat increases, since the structure formation process is less advanced and, therefore, overdensities are smaller. Future observational efforts should focus on charting this parameter space. In particular, we predict that the identification of sightlines with $\taulos \sim 2$--$3$ and $\Sigma_\mathrm{gal} (d < 10 \, \mathrm{Mpc}/h) / \langle \Sigma_\mathrm{gal} \rangle \lesssim 0.5$ would be ideal to more firmly pinpointing the reionization history of the Universe.

\begin{figure}
    \includegraphics[width=0.99\columnwidth]{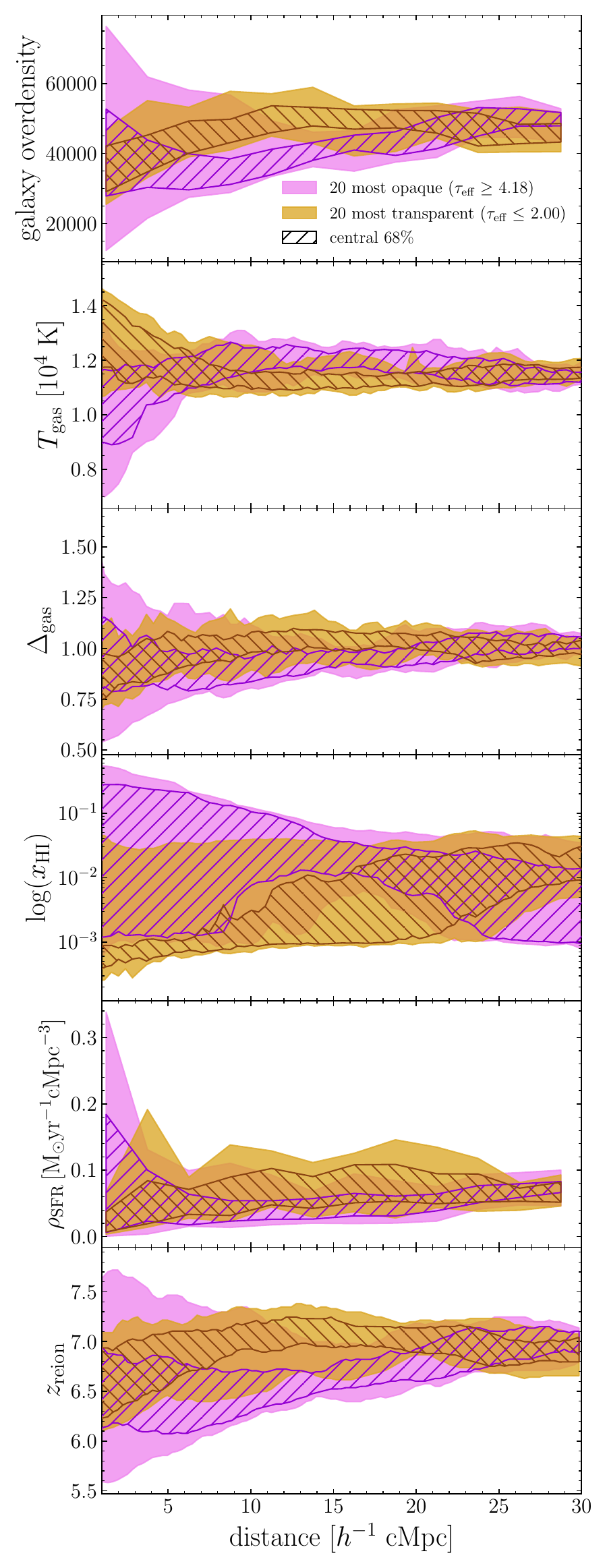}
    \caption{Radial profile of (from top to bottom, respectively) galaxy overdensity, gas temperature, gas overdensity, \HI fraction, \rev{density of star formation} and reionization redshift (see text for definition) around he 20 most opaque (purple shading and hatching) and most transparent (orange shading and hatching). The shaded regions show the envelope of all individual profiles, while the hatched ones report the central 68\% of data for each of them.    
    }
    \label{fig:physical_properties_los}
\end{figure}

\subsection{Physical properties around sightlines}
\label{sec:physical_properties}
Reassured by the overall agreement between our simulated lines of sight and available observations, in this Section we explore the predicted physical properties along and around the simulated sightlines, with particular focus on the difference between transparent and opaque LOS.

In Fig.~\ref{fig:physical_properties_los} we show (from top to bottom) the mean galaxy overdensity, gas temperature, gas overdensity, mass-weighted \HI fraction\footnote{We employ mass-weighted \HI fractions here rather than volume-weighted ones because of practical reasons. The \thesan simulations store the mass-weighted \HI fraction in so-called Cartesian outputs, which are much easier to analyze than the traditional snapshot format. Since we are interested in comparison between different groups of sightlines rather than absolute values, we expect no substantial difference when using mass- or volume-weighted \HI fractions, and therefore choose to favor the simplicity of analysis by choosing the former.}, 
density of star formation and median reionization redshift as function of distance from the sightlines. 
The purple (orange) shaded region marks the envelope of the distributions around the 20 most opaque (transparent) lines of sight in the simulation, while the hatched regions show the central 68\% of the data for each distribution. 
Notice that the number of sightlines used here is much smaller than previously used (\eg we used 100 in Fig.~\ref{fig:ngal_R_visual}). This ensures a much better visual clarity. In fact, the trends found are visible also when selecting the 100 most opaque and transparent sightlines, although their difference is smaller. 
The reionization redshift $z_\mathrm{reion}$ is defined for each cell of the \thesan Cartesian outputs as the largest redshift when a given cell is ionized to $x_\mathrm{HII} > 0.99$. 
Once again, we find that most of the differences are at $5 \lesssim d / [\hMpc] \lesssim 20$, where transparent sightlines have larger galaxy and gas density, lower temperature and lower \HI fraction. 

The aforementioned results paint a consistent picture of transparent sightlines probing regions where reionization occurred earlier and stronger (\ie the residual neutral fraction is lower) than average, due to the local abundance of galaxies. This is explicitly demonstrated in the bottom panel showing that at such scales $z_\mathrm{reion}$ is larger in transparent sightlines than in opaque ones. It also aligns well with the findings presented in Fig.~\ref{fig:ngt_corrcoeff_evol} and the relative discussion. 

Interestingly, at $d \lesssim 7 \, \hMpc$, the $z_\mathrm{reion}$ profile for opaque sightlines slightly turns over\footnote{To confirm the statistical significance of this, we have computed the global minimum position for each profile in the `most opaque' sample. Then, we performed a simple binomial test with null hypothesis that the global minima are equally probable at every radius. The p-value for such hypothesis is $0.04$, that is therefore rejected.}, indicating that on average galaxies close to the sightline ionize \textit{earlier} than those farther away. Together with the trend in galaxy density, which also turns over approximately at the same distance scale, this indicates that such opaque sightlines are probing an overdensity, which provides the photons that ionize the local IGM (and might shield them from incoming radiation). In other words, transparent sightlines probe regions that were ionized mostly outside-in, while opaque sightlines probe regions preferentially reionized inside-out. 

\rev{Finally, we record a small difference in the star formation rate density around different groups of sightlines (and, therefore, ionising photons production). This, however, is smaller than the difference in other properties investigated so far (the central 68\% of the two distributions overlap at all radial distances). Therefore, we conclude that the impact from different star formation activities is sub-dominant for the effects considered. We note, however, that the instantaneous star formation that we used is merely a proxy for the amount of ionizing photons produced over the lifetime of young stars in the galaxy (which typically emit ionising photons for $\sim 10$ Myr after birth, and up to $\sim 30$ Myr for some binary systems). Therefore, a more accurate account of the integrated stellar output might quantitatively change this result. However, we doubt that the change would be qualitative, since we obtain very similar results using the total stellar mass density (which is a proxy of the integrated star formation history galaxies around the sightline). We defer a more detailed investigation to a future work. }

We are now in the position to fully explain the puzzling finding in Fig.~\ref{fig:ngt_corrcoeff} that opaque sightlines appear to be most sensitive to closer galaxies than the full sample or the most transparent ones. Opaque sightlines are mostly affected by the overdensities reionizing them, and therefore are sensitive to smaller scales with respect to other sightlines (which sample preferentially underdense regions because they cover the majority of the volume) which are affected by radiation travelling (larger distances) through the underdense IGM, and therefore exhibit the characteristic scale dependence observed for the \galacc. 
This also explains the fact that opaque sightlines show a mildly negative (rather than positive) correlation of their opacity with the galaxy number density within few cMpc. Since they trace regions of inside-out reionization, a larger number of galaxies implies a stronger radiation field, that overcomes the negative impact of the local overdensity. 

Along (and very close to) the sightline, most of the physical quantities studied are identical in opaque and transparent LOS. The only difference found is that the range of (galaxy and gas) overdensity as well reionization redshifts probed is much larger in opaque sightlines, although the central value(s) of the distribution are very similar to those of transparent sightlines. 
This seems to indicate that within approximately $5\,\hMpc$ of the sightlines the IGM properties do not (strongly) influence its opacity (the difference in \HI fraction are a result of our selection, since not-strongly-ionized sightlines would not show up among the most transparent ones). 
The large difference in the scatter of the distribution of densities and $z_\mathrm{reion}$ at the smallest scales points to the fact that the volume of parameter space sampled is significantly smaller for the most transparent lines of sight with respect to the most opaque ones.

\begin{figure}
    \includegraphics[width=\columnwidth]{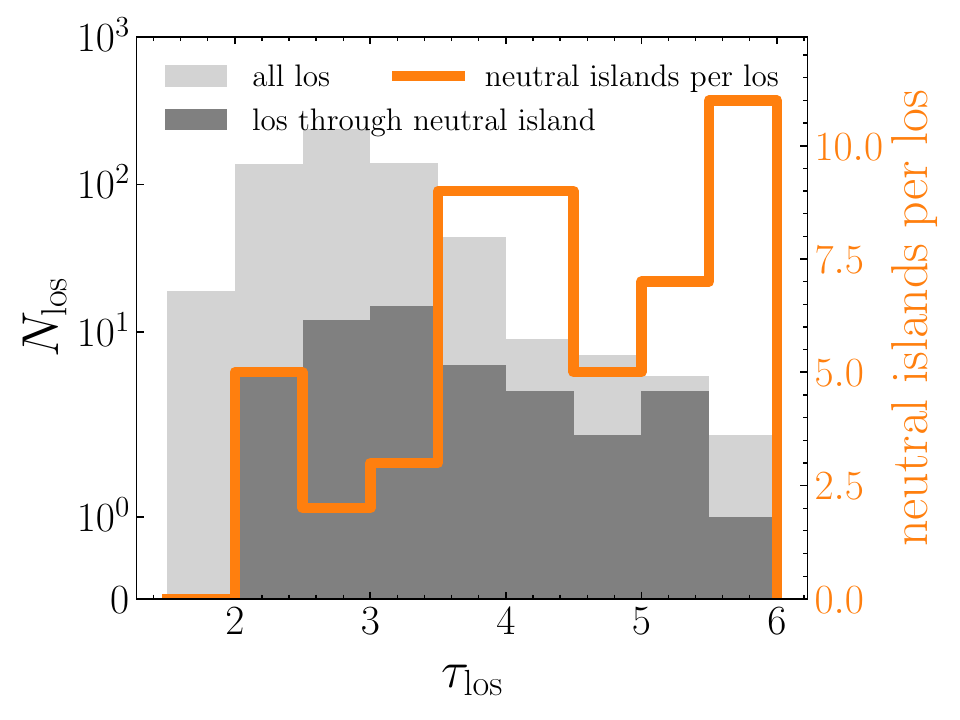}
    \caption{Distribution of optical depths ($\taulos$) in all sightlines (light gray histogram), in sightlines crossing a neutral island (dark gray histogram) and median number of neutral island crossed as function of $\taulos$ (orange histogram, referring to the right-hand-side vertical axis). Neutral islands promote the existence of high-$\taulos$ sighlines but are not necessary for their existence.}
    \label{fig:neutral_islands}
\end{figure}

\subsubsection{The impact of residual neutral islands}
An important question debated in the literature \citep[\eg][]{Kulkarni2019, Keating+2020,Nasir&DAloisio2020} is whether the presence of residual neutral islands in the IGM is necessary to explain the observed optical depths. 
We contribute to this debate by showing in Fig.~\ref{fig:neutral_islands} the fraction of sightlines that traverse a neutral island. We define the latter as a region of the IGM having a volume-averaged neutral fraction $x_\mathrm{HI} \geq 0.99$. At $z=5.7$, the 0.3\% of the IGM in \thesanone is classified as neutral island following our definition. 

In Fig.~\ref{fig:neutral_islands} we show the distribution of $\taulos$ for all sightlines (light grey histogram), and for those that cross a neutral island (dark grey histogram)\footnote{We have checked that there is no qualitative difference when requiring the neutral island to cover at least 1\%, 5\% or 10\% of the sightline, although the number of sightlines decreases. Similarly, there is no qualitative difference when defining a neutral island as $x_\mathrm{HI} \geq 0.9$ (instead of $x_\mathrm{HI} \geq 0.99$).}. The figure shows clearly that sightlines crossing a neutral island do not have preferentially large $\taulos$, as a consequence of the very large length over which their effective optical depth is computed potentially offsetting the absorption within the neutral IGM. However, the incidence of neutral islands increases towards large $\taulos$, reaching approximately 60\% for sightlines with $\taulos \gtrsim 4$. The fact that this never reaches 100\% (despite the very generous association between LOS and neutral islands) is a \textit{crucial} finding, since it demonstrates that in fully-coupled radiation-hydrodynamical simulations sightlines with large optical depths can be produced even in the absence of neutral islands. 

\rev{Additionally we check the median number of neutral patches crossed by LOS in a given $\taulos$ bin, which we show in the figure using an orange histogram (the vertical values can be read on the right-hand-side vertical axis). We also computed the length of such patches, obtaining median values as function of LOS opacity between $0.2$ and $0.6$ cMpc. The fact that opaque sightlines cross multiple small neutral regions, instead of having the optical depth dominated by a single extended neutral island, reinforces the conclusion that at the tail-end of reionization in the \thesan simulation we do not require extended neutral islands to produce very opaque sightlines. }

\rev{As discussed already, whether this conclusion can be extended to much longer troughs and/or darker sightlines is unclear, as it would require much larger simulation boxes that we can currently not afford. }

\section{Results from the thesan physics variations}
\label{sec:results_variations}
In this Section we show how some key results discussed above are affected by the processes explored in the physics variation runs within the \thesan suite. All these simulations have a lower resolution (by a factor of 8 in mass) than \thesanone. In particular, we employ the following simulations:
\begin{itemize}
    \item \thesanhigh, where ionising photons are produced only by galaxies residing in haloes with total mass $M_\mathrm{halo} \geq 10^{10}\,\Msun$.
    \item \thesanlow, where ionising photons are produced only by galaxies residing in haloes with total mass $M_\mathrm{halo} \leq 10^{10}\,\Msun$.
    \item \thesansdao, where cold dark matter is replaced by the strong Dark Acoustic Oscillation model.
    \item \thesanwc, where the stellar escape fraction is recalibrated to approximately match the reionization history of \thesanone. 
\end{itemize}
Each model has a different reionization history. In order to factor out this difference, in the following we present results at the redshift where their volume-averaged \HII fraction is the closest to the one in \thesanone at $z=5.7$. These redshift are: $z=6.73$ for \thesanlow, $z=5.58$ for \thesanhigh, $z=5.94$ for \thesansdao and $z=5.83$ for \thesanwc.

\begin{figure}
    \includegraphics[width=\columnwidth]{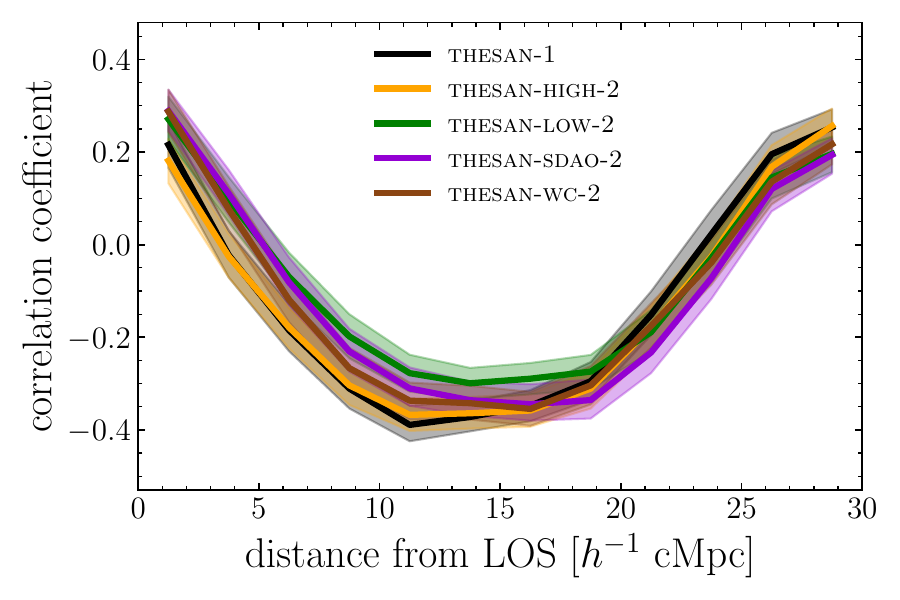}
    \caption{As Fig.~\ref{fig:ngt_corrcoeff} but showing only the curve relative to all galaxies (black in the original Figure) in the \rev{\thesanone}, \thesanhigh, \thesanlow, \thesansdao and \thesanwc runs.}
    \label{fig:ngt_corrcoeff_all}
\end{figure}

In Fig.~\ref{fig:ngt_corrcoeff_all} we show that, once we control for their different reionization histories, all models present a remarkably similar sensitivity of the sightline optical depth to galaxies at approximately $10 \lesssim d [\hMpc] \lesssim 20$. This aligns well with the findings in \citet{Thesan_igm} that the closely-related \galacc is robust against the explored model variations. The fact that we do not find a difference in this scale for \thesanhigh and \thesanlow, where reionization is sourced by very different galaxy populations, 
might appear worrying. However, as discussed in GB24, this proximity effect is not sourced by individual objects, that would therefore be sensitive to such difference in the sources, but rather by large ensembles of galaxies residing preferentially in overdense regions. In such configurations, the origin of individual photons loses importance in favour of the global photon output from the entire overdensity. Since we show results at similar $x_\mathrm{HII}$, the total photon output of these regions is also somewhat matched to each other. Therefore, the scale of influence becomes disconnected from the scale of the ionized bubbles, which differs in the \thesanhigh and \thesanlow models. 

The similarity in Fig.~\ref{fig:ngt_corrcoeff_all} is reflected by an equal similarity in the joint distribution of far-to-close overdensity ratio and line-of-sight optical depth (\ie the equivalent of Fig.~\ref{fig:oden_ratio_vs_tau}), which we omit for the sake of brevity. 

\begin{figure}
    \includegraphics[width=\columnwidth]{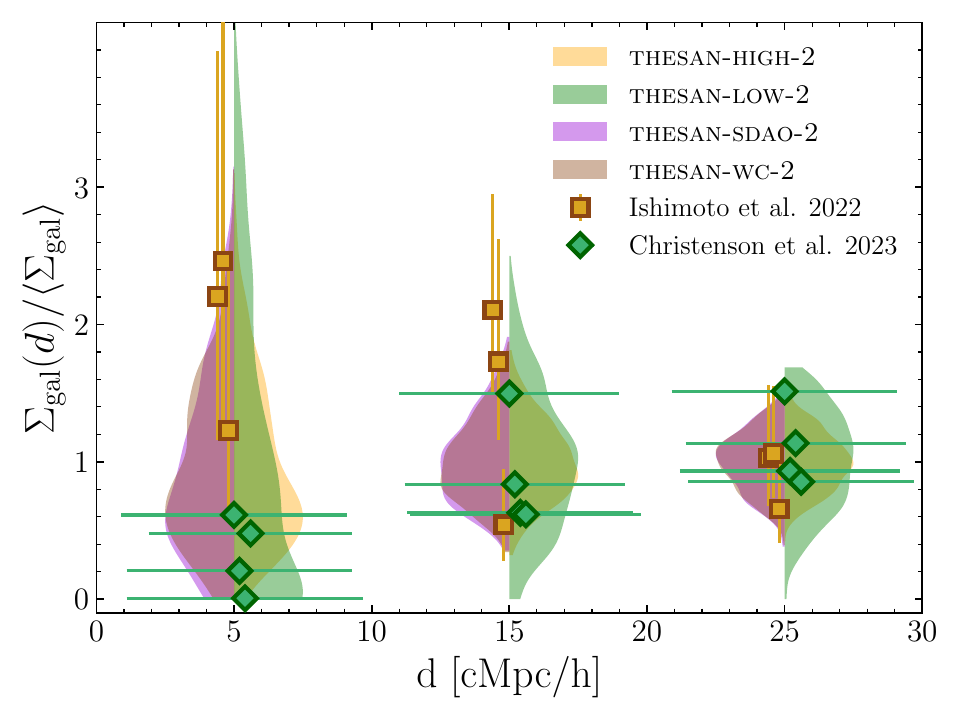}
    \caption{As Fig.~\ref{fig:oden_vs_dist_obs} but showing results for the \thesanhigh and \thesanlow on the right side of each distance bin and for \thesansdao and \thesanwc on the left side. For visual clarity, we only present the distributions computed using all the lines of sight (corresponding to the red left-side violin in Fig.~\ref{fig:oden_vs_dist_obs}). }
    \label{fig:sigma_dist_all}
\end{figure}

\begin{figure}
    \includegraphics[width=\columnwidth]{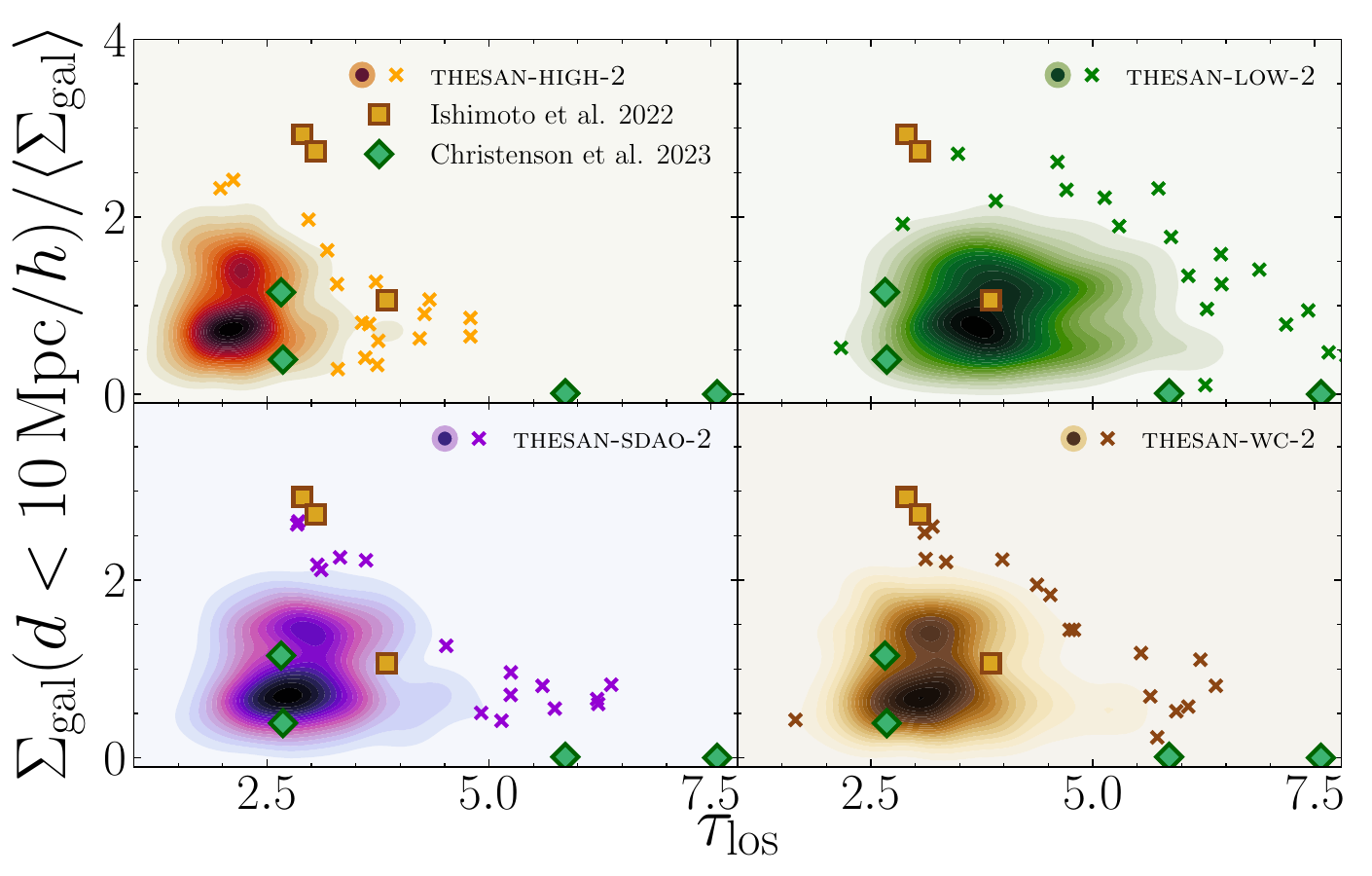}
    \caption{As Fig.~\ref{fig:oden_vs_tau_obs} but showing results for the \thesanhigh, \thesanlow, \thesansdao and \thesanwc runs. }
    \label{fig:oden_tau_all}
\end{figure}

We compare the different physics variations with observations in Fig.~\ref{fig:sigma_dist_all} and Fig.~\ref{fig:oden_tau_all}. As we have done for \thesanone (see Sec.~\ref{subsec:comparison_with_obs}), we separately match the number density of galaxies used in each model to the observed one by considering only galaxies more massive than a mass threshold. All models results in similar thresholds, except for \thesanlow, which demands a lower value. This stems from the \HII fraction-matching approach aforementioned, resulting in a significantly higher redshift for \thesanlow than for the other runs, and therefore in structure formation at an earlier stage. 

In Fig.~\ref{fig:sigma_dist_all} we present the distribution of galaxy overdensity in three different distance bins from all the sightlines in each model. For visual clarity, we only show the distribution computed for all sightlines and place two models on each side of the bin center. The distributions appear very similar to each other, with the exception of \thesanlow showing a broader range of overdensities at all distances. This is a consequence of the lower mass threshold used in this model to match the observed LAEs density, therefore increasing the likelihood of finding large galaxy overdensities.

In Fig.~\ref{fig:oden_tau_all}, we show the 2-dimensional distribution of $\taulos$ and galaxy overdensity within $10 \hMpc$, alongside the observations of \citet[][yellow squares]{Ishimoto+2022} and \citet[][green diamonds]{Christenson+2023}. All models present somewhat similar distributions (coloured contours), but some clear differences can be seen. In \thesanhigh (top left) the distribution is shifted towards more transparent sightlines, as a consequence of the fact that its \HII fraction-matching redshift is the lowest, and therefore underdense regions are more common and more underdense, promoting the \Lya transmission. For the same reason, the distribution in \thesanlow reached significantly larger optical depths than the other ones. 
In fact, \thesanlow features some sightlines as opaque as the most-opaque LOS known (the right-most diamond in the Figure) and more easily reproduces the observations of \citet{Christenson+2023}. 

All models struggle to reproduce the large overdensities around the sightlines observed by \citet{Ishimoto+2022}, although there is somewhat less tension in the \thesansdao and \thesanwc models. This is a consequence of the fact that these models are analysed at a slightly higher redshift than \thesanone or \thesanhigh, and therefore the observed number density of LAEs is matched with a slightly lower mass threshold. This, in turn, increases the probability of finding many such galaxies close to each other. 
Interestingly, \thesanlow, which uses a much smaller mass threshold than any other run analysed to match the observed density of LAEs, also struggles to explain the observations by \citet{Ishimoto+2022}. In fact, despite showing overall more sightlines with large overdensities, their optical depth is too large. Overall, \thesanlow performs worse than \thesansdao or \thesanwc in this comparison. 
We caution against using this as an evidence against a model of reionization driven by low-mass galaxies. As described in \citet{Thesan_igm}, reionization in \thesanlow completes much earlier than indicated by observations as a consequence of a sub-optimal choice of stellar escape fraction. Therefore, results from such run should be considered suggestive but not quantitatively sound. 
Nevertheless, these differences underscore the usefulness of observations like the \taun relation, which combine information on the reionization state of the IGM and the process of structure formation.

\section{Summary and Conclusions}
\label{sec:conclusions}
The results presented in the previous two Sections demonstrate the intimate connection between the \galacc and the \taun relation, clarifying how the insights obtained on the former can shed light on the observed features of the latter. They also demonstrate the importance of state-of-the-art radiation-hydrodynamical simulations like \thesan that combine accurate galaxy formation and IGM physics, as well as the added value brought by studies of physics variations. 

Our main findings are:
\begin{itemize}
	\item The LOS effective optical depth ($\taulos$) shows the strongest {(anti-)} correlation with the abundance of galaxies at distances $d \sim 15 \, \hMpc$. This is reduced to $d \sim 7 \, \hMpc$ for the most opaque sightlines as a result of the fact that reionization is less developed around them and therefore the nearby (highly-) ionized bubbles are smaller. 
	\item The \taun relation evolves with redshifts, tracking the progressive reionization of the IGM and the concurring expansion of (highly-)ionized bubbles. In particular, the galaxies most relevant for the sightline optical depth are closer at higher redshift, and are essentially \textit{along} the sightline at $z \gtrsim 6.5$. 
	\item The most transparent sightlines preferentially have larger galaxy and gas densities, lower temperature,  lower \HI fraction and earlier median reionization redshifts at radial distances $5 \lesssim d/[\hMpc] \lesssim 20$ with respect to the most opaque ones. The profile of $z_\mathrm{reion}$ \textit{decreases} with increasing distance within approximately $7\,\hMpc$ of opaque sightlines, suggesting they probe overdense region driving the local reionization. The emerging physical picture is that transparent sightlines probe regions that were ionized mostly outside-in, while opaque sightlines probe regions preferentially reionized inside-out. 
	\item Along and very close to the sightline, there is no difference between opaque and transparent LOS, except for a larger scatter in the (galaxy and gas) overdensity and reionization redshift along opaque sightlines.
	\item The \thesanone simulation is able to reproduce well the salient observed features of the \taun relation, although it struggles at matching the most extreme datapoints, likely as a result of its somewhat limited volume.
	\item Only approximately half of the sightlines with large optical depth ($\taulos \gtrsim 4$) at the tail-end of reionization ($z=5.7$, $x_\mathrm{HI} \sim 10^{-3}$) cross a neutral island. \rev{Even when they do, such neutral patches are small and numerous (median length between $0.2$ and $0.6$ cMpc and median number per sightline between 2.5 and 10, depending on the optical depth).} Therefore, we predict that the observation of such opaque sightlines does not automatically imply the existence of neutral islands. \rev{Nevertheless, it should be noted that our limited box size prevents us from probing the most opaque lines of sight observed, therefore leaving open the question whether such extreme cases can be reproduced without extended regions of neutral gas in the IGM.}
    \item Once the different reionization histories are accounted for, all the physics variations investigated show an identical radial sensitivity to the galaxy number density, \ie they all are most sensitive to galaxies at distances $d \sim 10 \, \hMpc$
	\item All the physics variations investigated reproduce to a similar degree the observed galaxy overdensity radial profile and struggle to match the joint distribution of local galaxy overdensity and LOS optical depth. The model performing the worst is \thesanhigh, where reionization is accomplished by large ($M_\mathrm{halo} \geq 10^{10} \, \Msun$) galaxies only.
\end{itemize}

The implications of our findings are profound. Firstly, they provide a clear physical picture of the galaxy-IGM connection at the end of reionization. This helps interpreting the still-scarce observations of the \taun relation. Second, we provide compelling evidence that neutral islands are not \textit{necessarily} required to explain the large optical depths observed at the tail end of reionization, although in our model their existence explains approximately 50\% of the largest-$\taulos$ sightlines. Third, our results highlight the potential of observations of the \taun relation, which combine measurements of cosmic reionization and structure formation, providing a crucial connection between the two. 

While ours results are encouraging, the extreme scarcity of observations does not allow us to draw robust conclusions. The situation calls for an effort to increase the number of observed sightlines (ideally by at least an order of magnitude) and, simultaneously, probe different redshifts, since we predict a strong time evolution of this relation, that carries information on the process of reionization. Ultimately, our work represents another step forward in the important quest to unveil the connection between first galaxies and reionization, that not only will test our understanding of the first billion years of the Universe, but promises to be central in the study of structure formation in the upcoming decade.

\section*{Acknowledgements}
We are thankful to the community developing and maintaining software packages extensively used in our work, namely: \texttt{matplotlib} \citep{matplotlib}, \texttt{numpy} \citep{numpy}, \texttt{scipy} \citep{scipy}, \texttt{cmasher} \citep{cmasher} and \texttt{CoReCon} \citep{corecon}.

\section*{Data Availability}
All simulation data and post-processing data products are publicly available at \url{www.thesan-project.com} and thoroughly described in \citet{Thesan_data}.

\section*{Author contributions}
We list here the authors contribution following the CRediT\footnote{\url{https://www.elsevier.com/researcher/author/policies-and-guidelines/credit-author-statement}} system. 
EG: conceptualization, methodology, software, formal analysis, validation, writing -- original draft, writing -- review and editing, visualization, supervision, project administration. 
VB: software, formal analysis, writing -- review and editing.
AS: writing -- review and editing.

\bibliographystyle{mnras}
\bibliography{bibliography}

\vspace{5cm}
\appendix
\section{Impact of the length of the spectra used in the analysis}
\label{app:different_spectral_lengths}

\begin{figure}
    \includegraphics[width=\columnwidth]{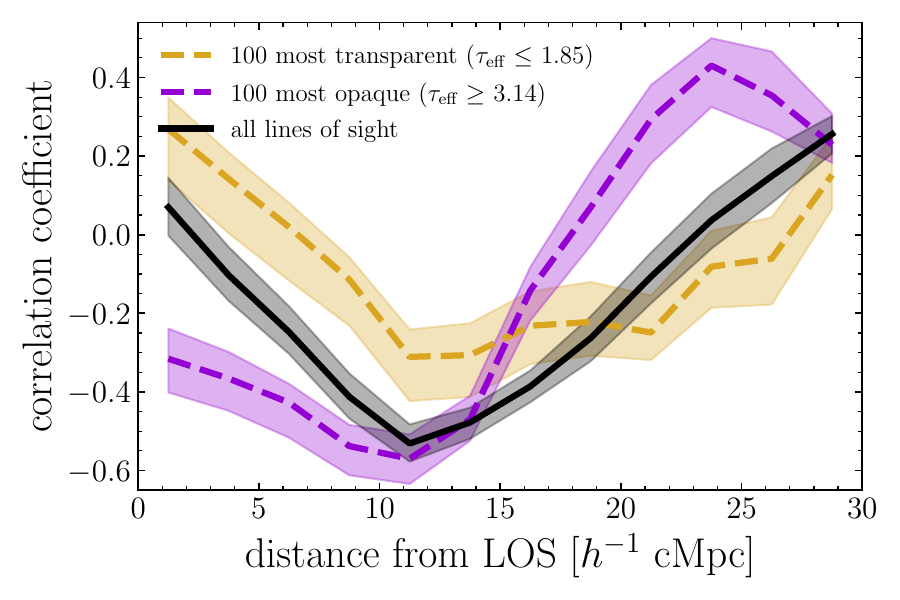}
    \caption{Same as Fig.~\ref{fig:ngt_corrcoeff} but now matching exactly the observational configuration, \ie $L_\mathrm{los} = 50$ Mpc and $L_\mathrm{gal} = 28$ Mpc.}
    \label{fig:ngt_corrcoeff_50_28}
\end{figure}

\begin{figure}
    \includegraphics[width=\columnwidth]{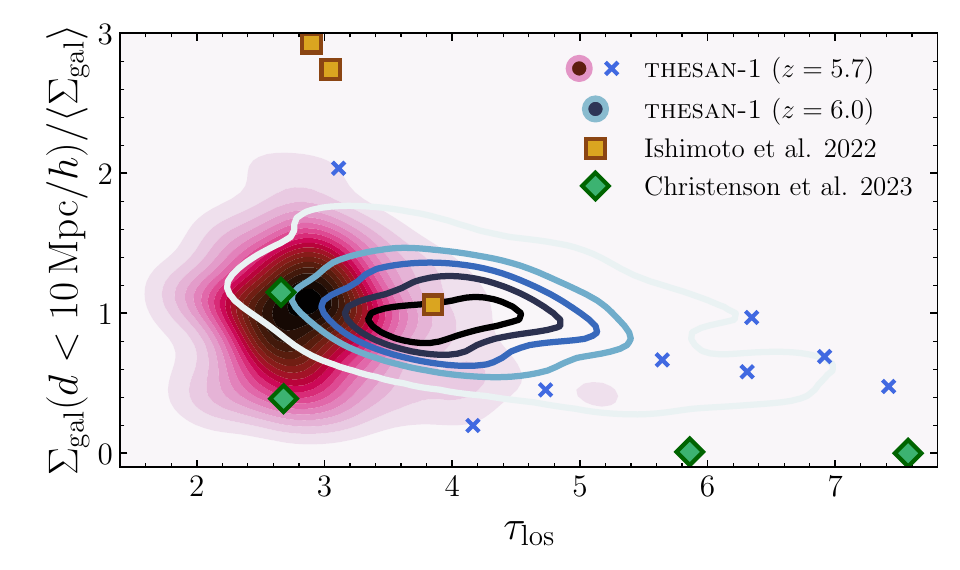}
    \caption{Same as Fig.~\ref{fig:oden_ratio_vs_tau} but now matching exactly the observational configuration, \ie $L_\mathrm{los} = 50$ Mpc and $L_\mathrm{gal} = 28$ Mpc.}
    \label{fig:oden_ratio_vs_tau_50_28}
\end{figure}

In this appendix we show the impact of varying the length of the sightline ($L_\mathrm{los}$) and of the region where galaxies are identified  ($L_\mathrm{gal}$, always assumed to be centered on the center of the sightline). In particular, throughout the paper we have chosen $L_\mathrm{los} = L_\mathrm{gal} = 50$ Mpc for a practical reason (\ie our pre-computed sightlines are $100 \, \hMpc$ long, so we can split all of them in half and double their number). In this Appendix, we show that this choice bears no consequences when compared to the observational approach followed by \citet{Christenson+2023} and \citet{Ishimoto+2022}, namely $L_\mathrm{los} = 50 \, \hMpc$ (notice the factor $1/h$) and $L_\mathrm{gal} = 28 \, \hMpc$ (we assume the FHWM of the NB816 filter used in both works to identify LAEs to be representative of $L_\mathrm{gal}$). We note that this value of $L_\mathrm{los}$ implies that we can only make use of $300$ sightlines instead of the $600$ used throughout the paper.

As representative examples, we 
reproduce Fig.~\ref{fig:ngt_corrcoeff} and Fig.~\ref{fig:oden_ratio_vs_tau} using the new configuration in Fig.~\ref{fig:ngt_corrcoeff_50_28} and in Fig.~\ref{fig:oden_ratio_vs_tau_50_28}, respectively. 
Beyond some minor quantitative differences due to the different number of sightlines, the results are remarkably similar to the one presented in the main text. We therefore conclude that the analysis performed with $L_\mathrm{los} = L_\mathrm{gal} = 50$ Mpc can be faithfully compared to observations. 

\vspace{1cm}
\label{lastpage}
\end{document}